\documentclass{SciPost}
\usepackage{braket}
\usepackage{amsmath}
\usepackage[T1]{fontenc}
\usepackage{graphicx} % Required for inserting images
\usepackage{lipsum}
\usepackage{hyperref}
\usepackage{float}

\usepackage[bitstream-charter]{mathdesign}
\DeclareSymbolFont{usualmathcal}{OMS}{cmsy}{m}{n}
\DeclareSymbolFontAlphabet{\mathcal}{usualmathcal}

\fancypagestyle{SPstyle}{
\fancyhf{}
\lhead{\colorbox{scipostblue}{\bf \color{white} ~SciPost Physics Core }}
\rhead{{\bf \color{scipostdeepblue} ~Submission }}

\fancyfoot[C]{\textbf{\thepage}}
}

\hypersetup{
    colorlinks,
    linkcolor={red!50!black},
    citecolor={blue!50!black},
    urlcolor={blue!80!black}
}

\usepackage[bitstream-charter]{mathdesign}
\DeclareSymbolFont{usualmathcal}{OMS}{cmsy}{m}{n}
\DeclareSymbolFontAlphabet{\mathcal}{usualmathcal}

\fancypagestyle{SPstyle}{
\fancyhf{}
\lhead{\colorbox{scipostblue}{\bf \color{white} ~SciPost Physics Core }}
\rhead{{\bf \color{scipostdeepblue} ~Submission }}

\fancyfoot[C]{\textbf{\thepage}}
}

\begin{document}

\pagestyle{SPstyle}

\begin{center}{\Large \textbf{\color{scipostdeepblue}{
%%%%%%%%%% TODO: Write your article's title here
Evolution of the Berry curvature dipole in uniaxially strained bilayer graphene\\
}}}\end{center}

\begin{center}\textbf{
% Use (full) first name (+ middle name initials) + surname format.
% Separate subsequent authors by a comma, omit comma and use "and" for the last author.
% Mark the corresponding author(s) with a superscript symbol in this order
% \star, \dagger, \ddagger, \circ, \S, \P, \parallel, ...
Karel Cuypers\textsuperscript{$\star$},
Robin Smeyers, %\textsuperscript{1},
Bert Jorissen and %\textsuperscript{1} and
Lucian Covaci \textsuperscript{$\dagger$}
}\end{center}

\begin{center}
Department of Physics, NANOlight Center of Excellence, University of Antwerp, Belgium
%%%%%%%%%% END TODO: AFFILIATIONS
% Provide email address of corresponding author(s)
\\[\baselineskip]
$\star$ \href{mailto:karel.cuypers@uantwerpen.be}{\small karel.cuypers@uantwerpen.be}\,,\quad
$\dagger$ \href{mailto:lucian.covaci@uantwerpen.be}{\small lucian.covaci@uantwerpen.be}
\end{center}

\section*{\color{scipostdeepblue}{Abstract}}
\textbf{\boldmath{%
While in pristine bilayer graphene the Berry curvature dipole (BCD), a necessary ingredient for the nonlinear anomalous Hall effect, is zero, uniaxial strain can give rise to finite BCD. We investigate this by using a tight-binding (TB) approach build on the Slater–Koster parameterization to capture lattice deformation effects often missed by continuum models. We demonstrate that the BCD's evolution with strain and doping is highly sensitive to the choice in parameterization, particularly when including the longer range interlayer skew hoppings. Additionally, out-of-plane compression enhances the response by broadening the Dirac cones. These findings benchmark low-energy continuum models and highlight the necessity of realistic tight-binding models for accurately predicting strain-engineered Hall effects in bilayer graphene.
}}

\vspace{\baselineskip}

%%%%%%%%%% BLOCK: Copyright information
% This block will be filled during the proof stage, and finilized just before publication.
% It exists here only as a placeholder, and should not be modified by authors.
%\noindent\textcolor{white!90!black}{%
%\fbox{\parbox{0.975\linewidth}{%
%\textcolor{white!40!black}{\begin{tabular}{lr}%
%  \begin{minipage}{0.6\textwidth}%
%    {\small Copyright attribution to authors. \newline
%    This work is a submission to SciPost Physics Core. \newline
%    License information to appear upon publication. \newline
%    Publication information to appear upon publication.}
%  \end{minipage} & \begin{minipage}{0.4\textwidth}
%    {\small Received Date \newline Accepted Date \newline Published Date}%
%  \end{minipage}
%\end{tabular}}
%}}
%}
%%%%%%%%%% BLOCK: Copyright information

% For convenience during refereeing we turn on line numbers:
%\linenumbers

\vspace{10pt}
\noindent\rule{\textwidth}{1pt}
\tableofcontents
\noindent\rule{\textwidth}{1pt}
\vspace{10pt}
%\setlength{\parindent}{0pt}
%\newcommand{\forceindent}{\leavevmode{\parindent=3em\indent}}
%\begin{document}
\newlength{\figwidth}
\setlength{\figwidth}{0.925\textwidth}

\section{Introduction}

Recently, significant attention has been given to anomalous Hall effects in graphene\cite{ho_zero-magnetic-field_2021} and other two-dimensional materials \cite{kang_nonlinear_2019, xiong_tunable_2025, wang_nonlinear_2025, qin_strain_2021, duan_giant_2022, he_graphene_2022}, . Unlike the conventional Hall effect, which requires an external magnetic field, these effects arise from the non-zero Berry curvature (BC) of the electronic bands, which acts as an effective magnetic field in momentum space \cite{xiao_berry_2010}. Two types of anomalous Hall responses are linked to non-zero Berry curvature: linear and nonlinear responses. The linear anomalous Hall current is proportional to the Berry curvature, whereas the nonlinear effect is linked to the presence of a non-vanishing Berry curvature dipole (BCD) \cite{bandyopadhyay_non-linear_2024}. In 2021, both linear and nonlinear Hall effect currents were observed in corrugated bilayer graphene \cite{ho_zero-magnetic-field_2021}, whereas previously such responses had only been observed in the Weyl semi-metal WTe$_2$ \cite{kang_nonlinear_2019}. While the BCD in pristine bilayer graphene vanishes, the nonlinear Hall effect can emerge when inversion symmetry is broken and the Dirac cones are tilted by corrugation or strain.

The anomalous Hall effect can be studied theoretically through simulations of the full electrical response, but such approaches are computationally intensive. A more amenable approach for homogeneous systems is to relate the BC and the BCD to the properties of the Bloch wave functions in k-space. These can be obtained by direct diagonalization of the system's Hamiltonian. Previous work, such as Ref. \cite{battilomo_berry_2019}, found that a non-vanishing BCD appears in uniaxially strained bilayer graphene as a consequence of Fermi surface warping. In that work, the BC and BCD were calculated using an effective low-energy continuum Hamiltonian, in which strain effects enters as an strain-induced gauge potential.

A different approach based on a tight-binding (TB) model can be used to calculate the energy dispersion and wave functions of the material. Importantly, the TB model of bilayer graphene offers some advantages over continuum models, especially when it comes to studying the effects of lattice deformations and strain. Strain is incorporated by explicitly deforming the lattice using a displacement field and then modifying the hopping amplitudes accordingly. % One then only has to solve the TB Hamiltonian of the strained system.

In this paper, we calculate the band structure, the BC, and the BCD of strained bilayer graphene using a full TB model with Slater-Koster parametrization that includes longer range interlayer hoppings, i.e., the so-called $\gamma_3$ and $\gamma_4$ skew hoppings \cite{mccann_electronic_2013}. We consider the effects of different types and magnitudes of homogeneous strain. We mainly focus on uniform uniaxial strain along either the zigzag (ZZ) or armchair (AC) directions. Our results are then compared to earlier results obtained from the continuum model in \textit{Battilomo et al. (2019)} \cite{battilomo_berry_2019}. We show how the BCD is dependent on the Fermi level and how it is affected by the Fermi surface warping modified by the strain. \textit{Battilomo et al. (2019)} neglected the $\gamma_4$ hopping term, which breaks the electron-hole symmetry at low energies \cite{mccann_electronic_2013}. This term can significantly modify the BCD depending on how the model is parameterized. We focus on two parameter sets: one obtained through fitting to experimental measurements of bilayer graphene \cite{kuzmenko_determination_2009} and one obtained by fitting the band structure of bulk graphite obtained using various other techniques \cite{mccann_electronic_2013}. Here we analyze the differences between these parameterizations when modeling the band structure and BCD of strained bernal bilayer graphene.

The final part of the paper explores the effects of changing the interlayer distance on the BCD. This out-of-plane strain is naturally included in the TB framework by adjusting the interlayer hoppings as the layer spacing changes. We therefore investigate the effect of compressing or expanding the interlayer distance, which can be achieved by external pressure, on the Dirac cone and the resulting nonlinear Hall response. Throughout the paper, a small gate-induced interlayer potential difference is applied to break inversion symmetry and generate finite BC, while strain warps the Fermi surface and controls the magnitude and sign of the BCD.

\begin{figure*}[t]
\centering
\includegraphics[width=\figwidth]{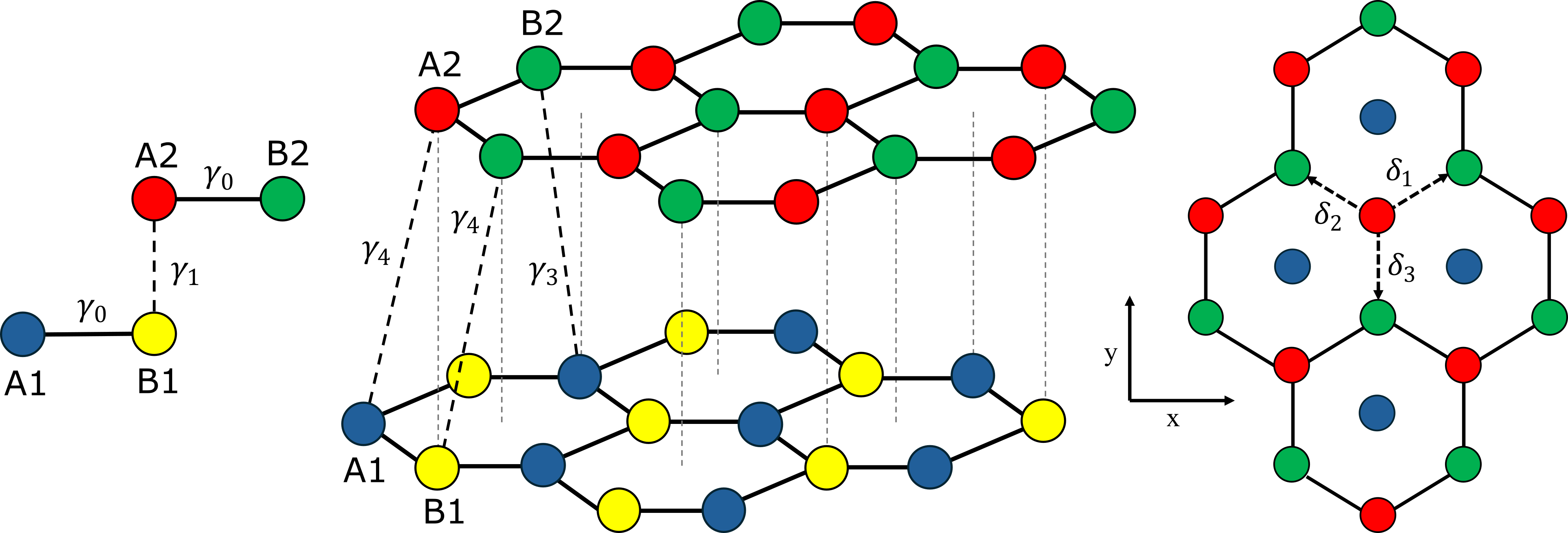}
\caption{\label{fig:TB_Bilayer_lattice} The tight-binding model of AB stacked bilayer graphene.}
\end{figure*}

\section{Methods}
% [76] is Malard L M, Nilsson J, Elias D C, Brant J C, Plentz F, Alves E S, Castro Neto A H and Pimenta M A 2007 Phys. Rev. B 76 201401(R) -- if you used this TB model mention it explicitly
\subsection{The tight-binding model}

The numerical methods used in this paper rely on diagonalizing the TB Hamiltonian of the bilayer graphene system and extracting the eigenvalues and eigenstates. The TB Hamiltonian used is based on the Slonczewski-Weiss-McClure model for bilayer graphene as described by \textit{McCann and Koshino (2013)} \cite{mccann_electronic_2013}. The Hamiltonian includes the nearest-neighbor hopping $\gamma_0$ as well as the interlayer perpendicular $\gamma_1$ and skew hoppings $\gamma_3$ and $\gamma_4$ (see Fig.\ref{fig:TB_Bilayer_lattice}). Values for these parameters can be obtained by fitting either measurements, such as infrared spectroscopy, or density functional theory calculations of the band structure \cite{kuzmenko_determination_2009, dresselhaus_intercalation_2002}.

When modeling bilayer graphene using the TB model, the inclusion of the $\gamma_3$ hopping term, between non-dimer sites (A1 and B2), modifies the low-energy quadratic dispersion, resulting in a central Dirac cone surrounded by three trigonally arranged satellite Dirac points, this is known as trigonal warping. Including the $\gamma_4$ hopping, between a non-dimer and dimer site (either A1 and A2 or B1 and B2) breaks electron-hole symmetry in the Dirac cones by shifting the satellite Dirac cones upward in energy \cite{mccann_electronic_2013}. The magnitude of this shift is proportional to the size of the $\gamma_4$ term and is therefore affected by the choice of parameterization. Following \textit{McCann and Koshino (2013)} \cite{mccann_electronic_2013}, the TB Hamiltonian in momentum space, in the basis $\{A1, B1, A2, B2\}$, reads:

\begin{equation}
    H(\vec{k}) = 
    \begin{bmatrix}
            \epsilon_{A1} & -\gamma_0 f(\vec{k}) & \gamma_4 f(\vec{k}) & -\gamma_3 f^*(\vec{k}) \\
            -\gamma_0 f^*(\vec{k}) & \epsilon_{B1} & \gamma_1 & \gamma_4 f(\vec{k}) \\
            \gamma_4 f^*(\vec{k}) & \gamma_1 & \epsilon_{A2} & -\gamma_0 f(\vec{k}) \\
            -\gamma_3 f(\vec{k}) & \gamma_4 f^*(\vec{k}) & -\gamma_0 f^*(\vec{k}) & \epsilon_{B2}
    \end{bmatrix},
\end{equation}

where:

\begin{equation}
    f(\vec{k}) = \sum^3_{l} e^{i\vec{k}\cdot \delta_l}.
\end{equation}

Here $\delta_l$ are the nearest-neighbor vectors $\delta_1 = a\left(\frac{\sqrt{3}}{2}, \frac{1}{2}\right)$, $\delta_2 = a\left(-\frac{\sqrt{3}}{2}, \frac{1}{2}\right)$ and $\delta_3 = - a\left(0, 1\right)$ with $a \approx 1.42$\AA being the carbon-carbon distance (see Fig.~\ref{fig:TB_Bilayer_lattice}). The $\epsilon_i$ model on-site potential terms, which can be used to define a gate-induced perpendicular electric field as an interlayer potential difference.

\subsection{Modeling strain}

Straining the lattice changes the interatomic distances. For uniform uniaxial strain, the nearest-neighbor vectors become:
\begin{align}
   \delta_1' &= a \left( \frac{\sqrt{3}}{2} (1 + \epsilon_{x}), \frac{1}{2} (1 + \epsilon_{y}) \right) \\
   \delta_2' &=  a \left( \frac{\sqrt{3}}{2} (-1 + \epsilon_{x}), \frac{1}{2} (1 + \epsilon_{y}) \right) \\
   \delta_3' &= a \left( 0, (-1 + \epsilon_{y}) \right)
\end{align}
where $\epsilon_{x}$ and $\epsilon_y$ are the components of the strain tensor in the $x$ (ZZ) or $y$ (AC) direction, respectively. For clarity,  we consider in this work that these two components are independent, i.e. we ignore the finite Poisson ratio of graphene. As the lattice is strained, the distance between atoms changes, affecting the hopping amplitudes. These are recalculated for each strained configuration based on Slater-Koster type functions \cite{slater_simplified_1954}. This approach was used in \textit{Moon, Koshino, 2012} \cite{moon_energy_2012} to model the intra- and interlayer hoppings for twisted bilayer graphene. We adopt a similar model for uniaxially strained bilayer graphene. The strain-dependent hopping between two $p_z$ orbitals separated by a vector $\vec{d}$ is:

\begin{equation}
    - \gamma_n(\vec{d}) = V_{pp\pi}(d) \left[ 1 - \left( \frac{\vec{d} \cdot \vec{e}_z }{d} \right)^2 \right] + V_{pp\sigma}(d) \left( \frac{\vec{d} \cdot \vec{e}_z }{d} \right)^2,
\end{equation}
with:
\begin{align}
    & V_{pp\pi}(d) = V^0_{pp\pi} \exp{\left( -\frac{d-a_0}{r_0} \right)}, \\
    & V_{pp\sigma}(d) = V^0_{pp\sigma} \exp{\left( -\frac{d-d_0}{r_0} \right),}
\end{align}
where $V^0_{pp\pi} = -\gamma_0$ and $V^0_{pp\sigma} = \gamma_1$ are the hoppings in the unstrained case between intralayer and interlayer nearest-neighbors, respectively. $a_0 = a_{cc} = 1.42$ \r{A} and $d_0 = 3.35$ \r{A} are the in- and out-of-plane atomic separations and $r_0 = 0.453$ \r{A} is a decay length.

Using this approach, $\gamma_3$ and $\gamma_4$ would be equal whenever the interlayer distances are equal, since the local atomic environments between the dimer and non-dimer sites are not taken into account. In reality,  $\gamma_3$ connects non-dimer sites, whereas $\gamma_4$ connects dimer to non-dimer sites. Therefore, these are not equivalent even if the distances are the same. To account for this, we will rescale the strain updated skew hoppings to preserve the unstrained ratio:

\begin{equation}
    \gamma'_n = \gamma_n \left( 1 + \frac{\gamma_n(\vec{d}) - \gamma_n(\vec{d}_0)}{\gamma_n(\vec{d}_0)} \right) \text{for} \  n = 3,4,
\end{equation}
where $\gamma_n(d)$ denotes the Slater-Koster estimate for the corresponding hopping at distance $d$ and $d_0$ is the unstrained distance.

\subsection{The effective low-energy Hamiltonian}

\begin{figure*}[t]
\centering
\includegraphics[width=\figwidth]{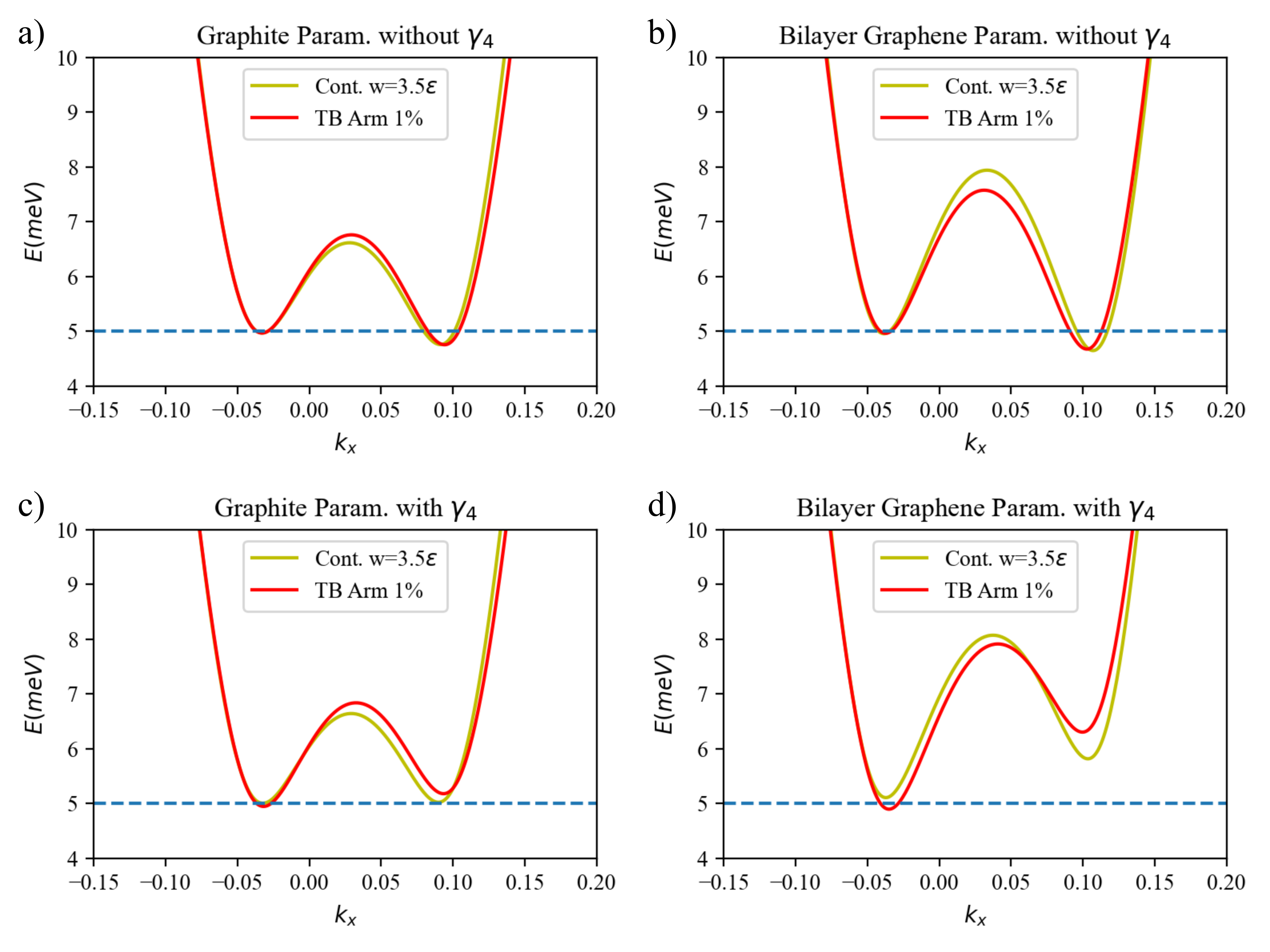}
\caption{\label{fig:TB_vs_analytical} The low-energy dispersion near the Dirac point sliced across the x-axis, e.g. $k_y=0$, calculated using both the TB and continuum models with the parameterizations given in Table \ref{tab:model_param} for bulk graphite a) without or b) with the $\gamma_4$ skew hopping and the band structure calculated with the parameterization for bilayer graphene c) without or d) with the $\gamma_4$ skew hopping. The dotted line indicates the value of the added gating potential, U = 10 meV, between the two layers.}
\end{figure*}

We compare our results to previous work on the continuum model described in \textit{McCann, Koshino, 2013} \cite{mccann_electronic_2013}. This model approximates the tight-binding Hamiltonian and describes an effective low-energy Hamiltonian that is valid for electron densities up to $n \approx 10^{13}$ cm$^{-2}$ \cite{mucha-kruczynski_landau_2011}. The analytical model also includes the effect of the $\gamma_3$ hopping term, which causes the trigonal warping, and the effects of the $\gamma_4$ hopping, which shifts the three-legged satellite Dirac cones up in energy. The effective low-energy Hamiltonian can be expressed as follows \cite{mccann_electronic_2013}:

\begin{equation}
\label{eq:anal_ham}
\begin{aligned}
    H =  & - \frac{1}{2m} \begin{bmatrix} 0 & (\pi^{\dagger})^2 \\ \pi^2 & 0\end{bmatrix} 
    + v_3 \begin{bmatrix} 0 & \pi \\ \pi^{\dagger} & 0\end{bmatrix}
    - \frac{v_3a}{4\sqrt{3} \hbar} \begin{bmatrix} 0 & (\pi^{\dagger})^2 \\ \pi^2 & 0\end{bmatrix}
    + \frac{2 v_0 v_4}{\gamma_1} \begin{bmatrix} \pi^{\dagger}\pi & 0 \\ 0 & \pi\pi^\dagger\end{bmatrix} \\
    & - \frac{U}{2} \left( \begin{bmatrix} 1 & 0 \\ 0 & -1 \end{bmatrix} 
    - \frac{2v^2}{\gamma_1^2} \begin{bmatrix} \pi^\dagger\pi & 0 \\ 0 & -\pi \pi^\dagger \end{bmatrix} \right)
    + \begin{bmatrix} 0 & w \\ w^* & 0\end{bmatrix}
\end{aligned}
\end{equation}
where $m=\frac{\gamma_1}{2v_0^2}$, $\pi = \xi q_x + iq_y$ with $(q_x,q_y)$ being the electron momentum relative to the Dirac point and $\xi = \pm 1$ the valley index, $m$ is the effective electron mass term and $v_n = \frac{\sqrt{3} a_0 \gamma_n}{2\hbar}$ is the Fermi velocity related to the hopping term $\gamma_n$. These parameters are related to the values of hopping amplitudes defined for the tight-binding model. The gating potential, $U=\frac{1}{2}(\epsilon_{A1}+\epsilon_{B1}-\epsilon_{A2}-\epsilon_{B2})$, is the applied potential difference between the two graphene layers. The last term $w$ is a gauge potential term, expressed in terms of two pseudo-gauge fields $A_0, A_3$\cite{vozmediano_gauge_2010}:

\begin{equation}
    w = A_3 - A_0 = \frac{3}{4} \gamma_0 (\eta_3 - \eta_0) (\delta - \delta') e^{-i 2 \theta},
\end{equation}
where $\eta_n = \frac{a}{\gamma_n} \frac{\partial \gamma_n}{\partial a}$ i.e. the in-plain change in the hopping parameter when the distance between carbon atoms changes, $\delta, \delta'$ are the eigenvalues of the strain tensor and $\theta$ is the angle between the strain direction and the principal strain axis \cite{mucha-kruczynski_landau_2011}. $w$ can be expressed in units of $\epsilon = mv_3^2/2$. When comparing with the TB band structure, Figure \ref{fig:TB_vs_analytical} shows that $w = 3.5 \epsilon$ is roughly equivalent to 1\% strain applied in the AC direction.

\subsection{The Berry curvature and Berry curvature dipole}

The BC is an intrinsic property of the band structure \cite{xiao_berry_2010}. It is associated with the change in the Berry phase of a quantum state during adiabatic expansion and is defined as the rotor of the Berry connection. The BC is defined as:
\begin{align} \label{eq:berry_curve}
    \vec{\Omega} (\vec{k}) = \Omega(\vec{k}) \vec{z}=\vec{\nabla}_{\vec{k}} \times \Braket{u_{\vec{k}} | -i \frac{\partial}{\partial \vec{k}} | u_{\vec{k}}},
\end{align}
where $u_k$ represents the Bloch functions of, in this case, bilayer graphene. For 2D systems, the BC is a $\vec{k}$ dependent vector pointing in the $\vec{z}$ direction.  The BC is both symmetric under inversion and anti-symmetric under time reversal. Therefore, it can only be non-zero when one of these symmetries is broken \cite{xiao_berry_2010}. In bilayer graphene, finite BC is typically achieved by breaking inversion symmetry with a gate-induced interlayer potential difference.

The intrinsic nonlinear Hall conductivity tensor is related to the BCD \cite{du_quantum_2021}: 

\begin{equation}
\chi^{in}_{abc}=\frac{e^3\tau}{4\hbar}(\epsilon_{abd}D_{cd} + \epsilon_{acd}D_{bd}),
\end{equation}
where $a,b,c=\{x,y,z\}$, $\tau$ is the relaxation time and the BCD is defined as
\begin{equation} \label{eq:berry_dipole}
    D_{ab} = \int_{BZ} f_0 \frac{\partial \Omega_b}{ \partial k_a},
\end{equation}
where in 2D the integral is $\int_{BZ} = \int d k_x dk_y / (2\pi)^2$, and $f_0$ is the Fermi-Dirac distribution. Here, the BCD is calculated at zero temperature by considering the integral to run over regions in the BZ with energy below the Fermi-level. The BCD can then be calculated by integrating the first derivative with respect to $k_a$ of the BC (also called the Berry dipole density or Berry moment) at every point with energy below the chosen Fermi level value. The BCD in pristine bilayer graphene vanishes due to the trigonal symmetry of the Dirac cone. The BCD around the whole Dirac cone cancels out \cite{battilomo_berry_2019} even if the Berry dipole density around each individual satellite cone is non-zero. The trigonal symmetry is broken by the applied uniaxial strain, thus giving rise to a finite total BCD.

The BC can be calculated numerically in a small square plaquette $\{\vec{k}_1 ... \vec{k}_4\}$  of area $\delta^2$ around each point in k-space. The wave function $u_{\vec{k}_i}$ is calculated in each corner point of the plaquette \cite{fukui_chern_2005}. The BC at grid point $\vec{k}$ is then calculated according to the following formula:
\begin{equation}
    \Omega_{\vec{k}} = \lim_{\delta \to 0} \frac{1}{\delta^2} \arg( \braket{u_{\vec{k}_1} | u_{\vec{k}_2}} \braket{u_{\vec{k}_2} | u_{\vec{k}_3}} \braket{u_{\vec{k}_3} | u_{\vec{k}_4}} \braket{u_{\vec{k}_4} | u_{\vec{k}_1}}).
\end{equation}

\subsection{Computational considerations}

All TB calculations in this paper were done using the open-source Python package: Pybinding \cite{moldovan_pybinding_2020}. The TB Hamiltonian is diagonalized using LAPACK routines \cite{virtanen_scipy_2020} to obtain the eigenvalues and eigenfunctions needed for the BC/BCD evaluations. The BC is calculated on a finite numerical grid. To ensure the calculation of the BC is accurate, the size of the plaquette around each grid point is much smaller than the spacing between the grid points, this was done to ensure convergence of the BCD values. Next, the derivative with respect to the finite grid is used to calculate the first moment of the BC. To calculate the BCD, the first moment of the BC is integrated numerically. The calculation of the BC can be computationally demanding and scales with the number of grid points. Thus, in order to accurately calculate the BCD, a denser grid is needed. To speedup the calculation we use a spline interpolation of the first moment of the BC to interpolate the values on a denser grid. This denser grid can then be used to calculate the numerical integral as a discrete sum and obtain a converged value for the BCD in a less computationally demanding manner. 

\section{Results}

\subsection{Effects of different parameterizations}
We consider two widely used parameterizations for the hopping amplitudes, obtained either by fitting to infrared spectroscopy of bilayer graphene \cite{kuzmenko_determination_2009} or to the band structure of bulk graphite \cite{dresselhaus_intercalation_2002}. Despite the latter describing bulk graphite, it is routinely used in the literature to parameterize bilayer graphene as well. In this paper, the band structure, BC, and BCD were calculated with both parameterizations to explore their impact.

\begin{table}[!h]
\centering
\begin{tabular}{c|c|c} 
    Parameter & Bilayer graphene (eV) & Graphite (eV) \\
    \hline
    $\gamma_0$ & 3.16 & 3.16 \\
    $\gamma_1$ & 0.381 & 0.38 \\
    $\gamma_3$ & 0.38 & 0.315 \\
    $\gamma_4$ & 0.14 & 0.044
\end{tabular}
\caption{Values of the hopping parameters obtained through fitting the TB model to experimental infrared spectroscopy results for bilayer graphene in \cite{kuzmenko_determination_2009} or through fitting to the band structure of bulk graphite \cite{dresselhaus_intercalation_2002}.}
\label{tab:model_param}
\end{table}

\begin{figure*}[t]
\centering
\includegraphics[width=\figwidth]{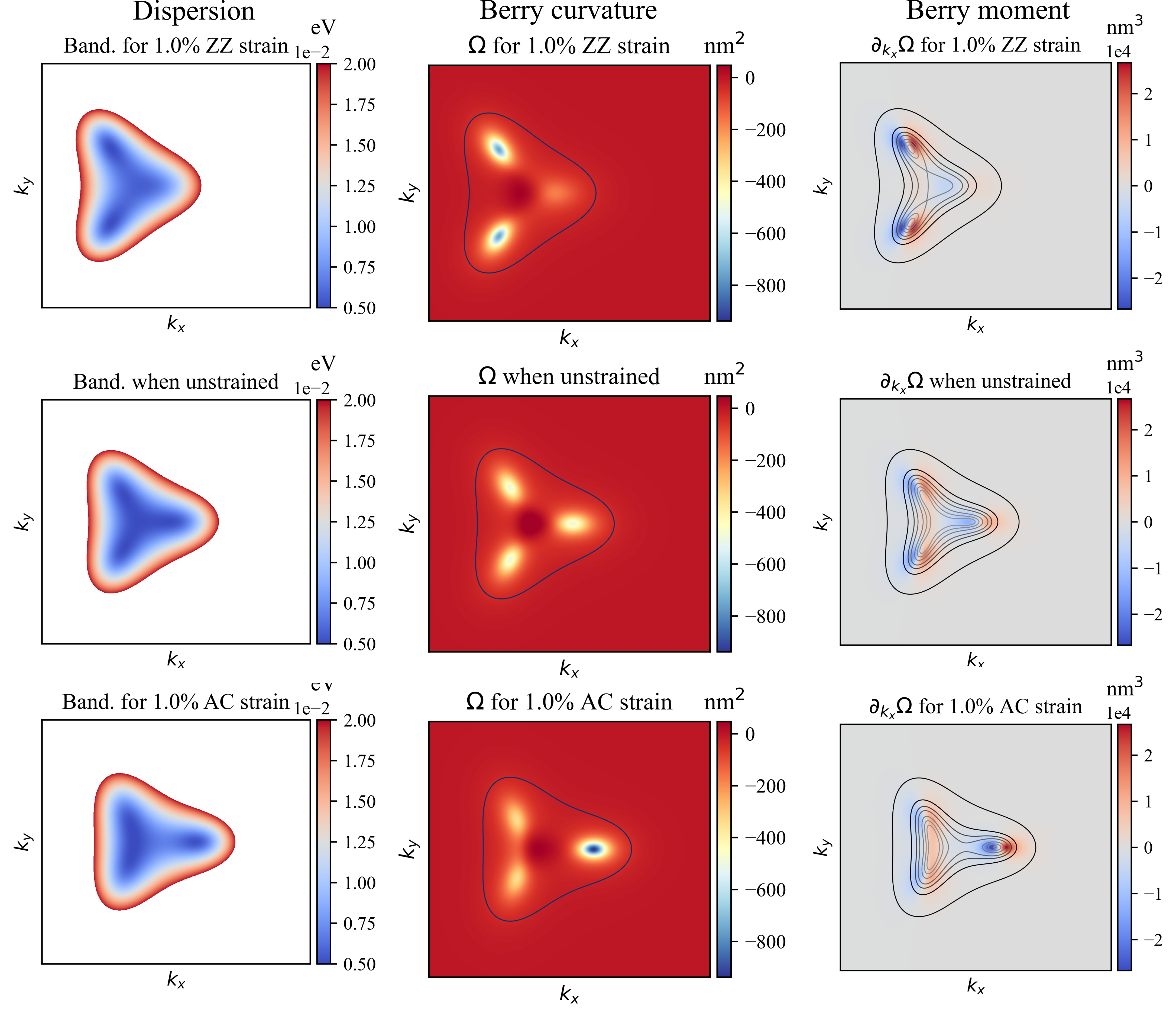}
\caption{\label{fig:ref67_berry} The electronic dispersion, the BC, and the first moment of the BC in the low energy region of the Dirac cone of bilayer graphene with an interlayer gating potential of U = 10 meV. The contour lines show the Fermi surface at various energies. These plots were made using the TB parameterization fitted to bulk graphite \cite{dresselhaus_intercalation_2002}.}
\end{figure*}

In the following, we compare the band structures calculated using the TB and analytical models with parameterizations given in table \ref{tab:model_param}. This is shown in figure \ref{fig:TB_vs_analytical}, which shows the dispersion along the x-axis for both models with roughly equivalent strains, with or without the $\gamma_4$ term. The $\gamma_4$ hopping term shifts the satellite Dirac cones, which are substantially larger for the bilayer-graphene parameter set, as $\gamma_4$ is larger in that case. The shift is large enough to significantly alter the shape of the Fermi surface at very low energies, which will affect the BCD.

\subsection{The warping of the Fermi-surface through strain}

The Dirac cone is located at the high symmetry K-point at the edge of the first Brillouin zone. When the lattice is strained, the K-point shifts in accordance with the lattice deformation. At small strains the Dirac cone will follow the position of the K-point \cite{pereira_tight-binding_2009, li_new_2014}. The position of the K-point on the x-axis in the presence of uniaxial strain is given by:
\begin{equation}
    \vec{K}^{\pm} = \pm \frac{4 \pi}{3 \sqrt{3a}} \left( 1 - \frac{\epsilon_x}{2} - \frac{\epsilon_y}{2}, \quad 0 \right).
\end{equation}

We now examine how strain affects the shape of the Dirac cone in the low-energy limit. The dispersion, BC, and BC dipole density of pristine and strained configurations (1\%) are shown in Figure \ref{fig:ref67_berry} for the bilayer fit. When strain is applied in the ZZ direction, the satellite Dirac cone located on the $k_x$ axis moves towards and eventually merges with the central cone. The merging of the cones will then give rise to a saddle point between the two remaining cones. When strain is applied in the AC direction, the satellite Dirac cones not located on the x-axis merge with the center Dirac cone. At the same time, the satellite Dirac cone located on the x-axis moves away from the central cone. These changes are a direct consequence of  breaking of the trigonal symmetry due to the applied uniaxial strain, resulting in significant changes in the BC dipole density at low energies.
%\begin{figure*}[h]
%\centering
%\includegraphics[width=\figwidth]{ref80_berry_plots_upto1.png}
%\caption{\label{fig:ref80_berry} The electronic dispersion, the BC, and the first moment of the BC in the low energy region of the Dirac cone in bilayer graphene with an added interlayer gating potential of 10meV. The contour lines show the warping of the Fermi surface at various energies. These plots were made using the bilayer-graphene TB parameterization.}
%\end{figure*}

Changes in the topology of the low-energy dispersion induced by strain have been described previously using the low-energy effective Hamiltonian in the continuum model \cite{mucha-kruczynski_landau_2011} shown in equation (\ref{eq:anal_ham}). This model incorporates strain as a gauge potential term added to the Hamiltonian. We can identify that a real positive value of the gauge term $w$ is equivalent to uniform uniaxial strain in the AC direction, while a real negative value is equivalent to a uniform uniaxial strain in the ZZ direction. In our TB model, the x-axis was chosen along the ZZ direction and the y-axis along the AC direction. 

\subsection{The Berry curvature dipole as a function of the electron density}
\begin{figure*}[t]
\centering
\includegraphics[width=\figwidth]{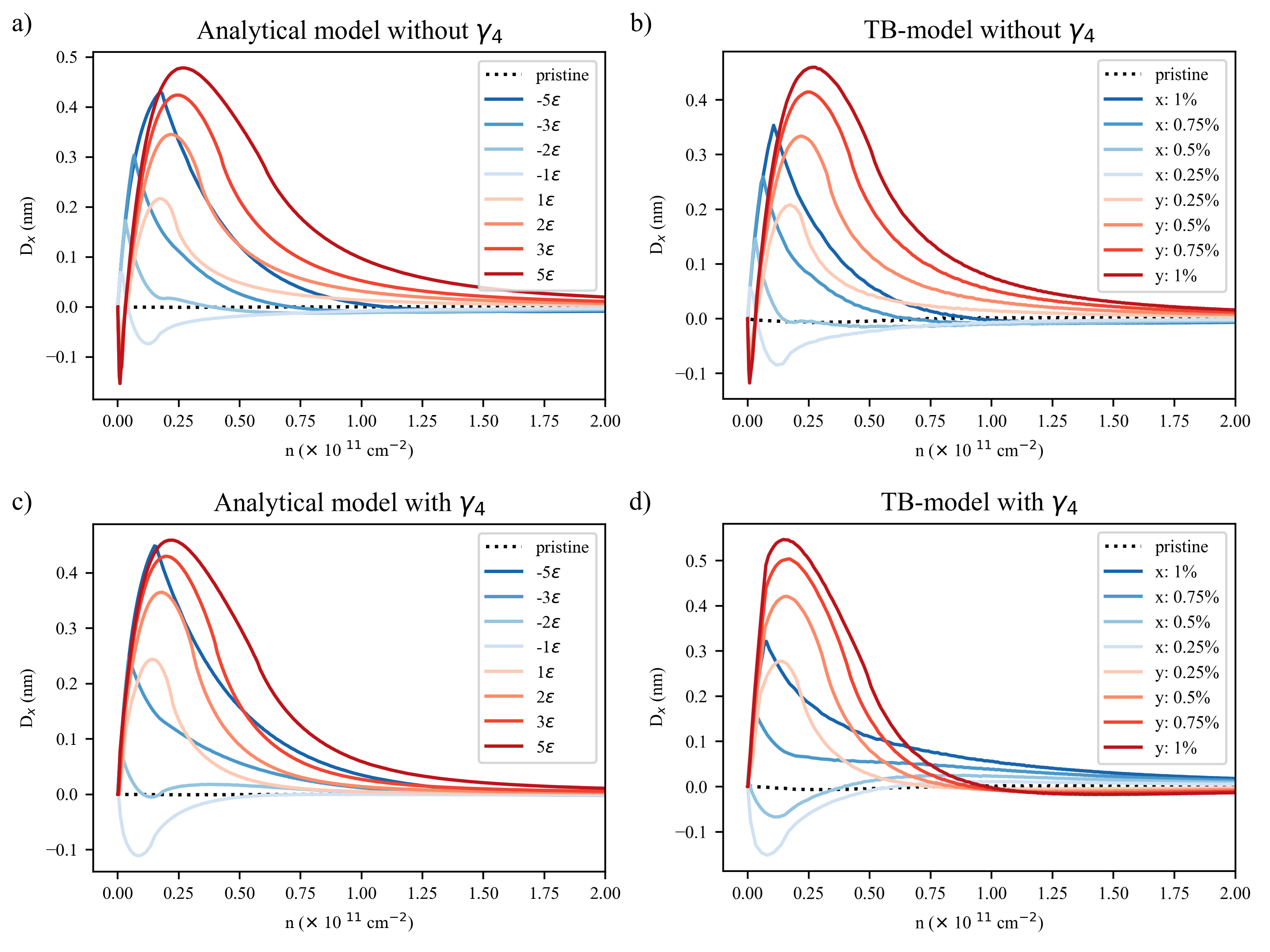}
\caption{\label{fig:ref67_dipole} The BCD as a function of the electron density calculated with the bulk graphite parametrization with U = 10 meV.}
\end{figure*}

\begin{figure*}[!bht]
\centering
\includegraphics[width=\figwidth]{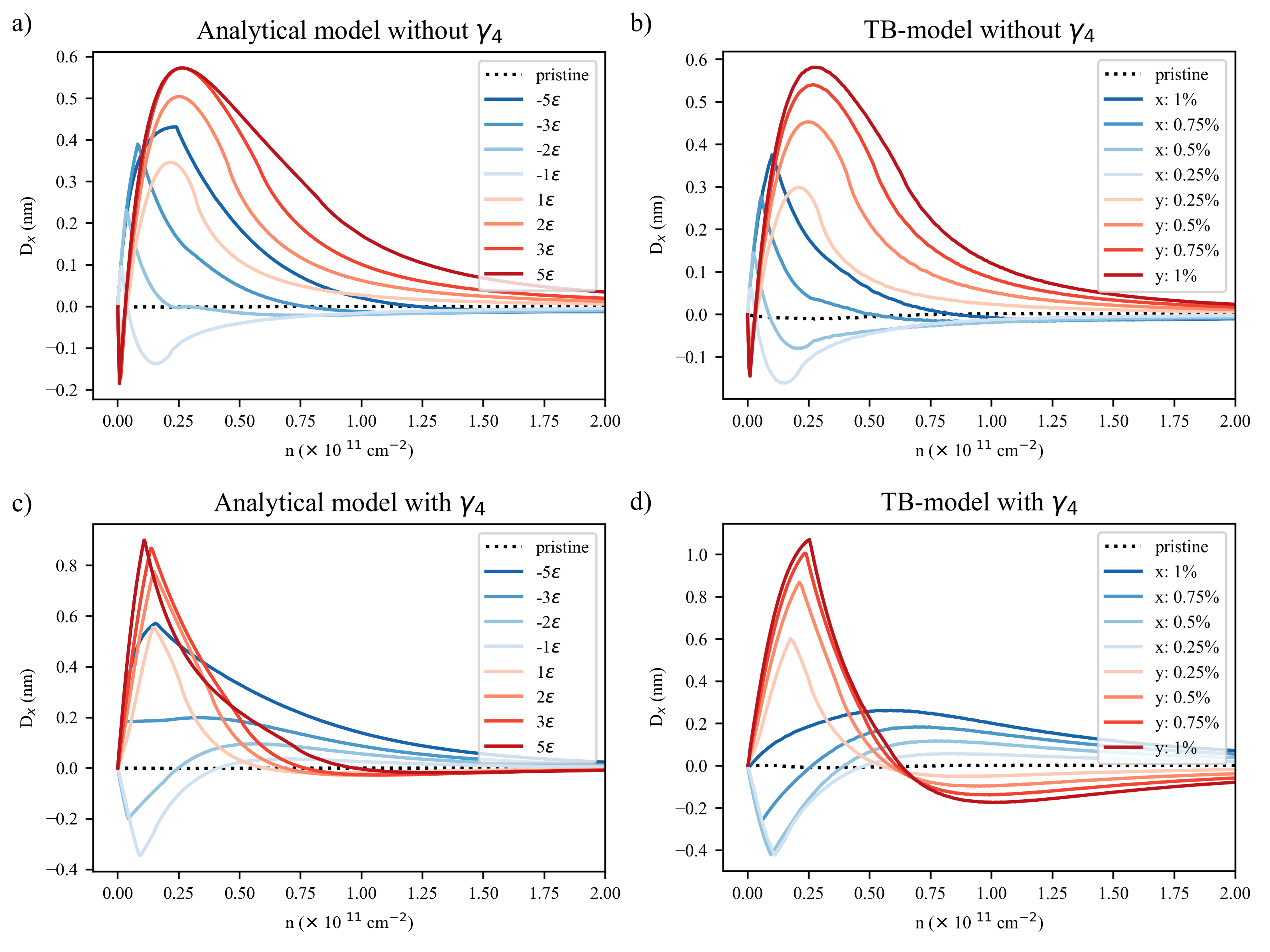}
\caption{\label{fig:ref80_dipole} The BCD as a function of the electron density calculated with the bilayer graphene parametrization with U = 10 meV.}
\end{figure*}

The BCD was calculated from the BC as a function of the electron density, according to  equation (\ref{eq:berry_dipole}). Figure \ref{fig:ref67_dipole} shows the BCD as a function of the electron density calculated using the parameterization for bulk graphite with both the TB and analytical continuum models. Figure \ref{fig:ref80_dipole} also shows the Berry curvature dipole as a function of the electron density, now calculated using the same models with the parameterization for bilayer graphene.

Figure \ref{fig:ref67_dipole} shows the Berry curvature calculated for the continuum and TB models. Panels \ref{fig:ref67_dipole}a and \ref{fig:ref67_dipole}b show the results for the models without the inclusion of the $\gamma_4$ skew hopping term, while panels \ref{fig:ref67_dipole}c and \ref{fig:ref67_dipole}d show the results with the $\gamma_4$-term. The $\gamma_4$-term was omitted in order to compare with earlier results from \textit{Battilomo et al. (2019)} \cite{battilomo_berry_2019}, where the Berry curvature dipole was calculated using the analytical model, in which the $\gamma_4$ term was also omitted. The results shown in Figure \ref{fig:ref67_dipole} for the analytical model without $\gamma_4$ agree well with the results in \textit{Battilomo et al. (2019)}. 

%In general, we find a good qualitative match for the BCD calculated with the TB and continuum models. The BCD will be negative for strain in the ZZ direction at low strain, and become positive and increase in size as strain increases to 1\%. For strain in the AC direction, the BCD will be positive and increase with strain up to 1\%.

The results shown in Figure~\ref{fig:ref67_dipole} can be compared to those shown in Figure~\ref{fig:ref80_dipole}. These figures show that the BCD as a function of the electron density or Fermi level, depends significantly on the chosen parameterization. Panels a) and b) of these figures show the results for the model where the $\gamma_4$-term was omitted. These results are very similar to those obtained using the parameterization fitted to bulk graphite. Panels c) and d), by contrast, show the results when the $\gamma_4$ term is included. In this case, notable differences emerge between the two parameterizations.

With $\gamma_4$ included, the BCD visibly changes sign at an electron density of $n \sim 7\times 10^{-12} cm^{-2}$ as seen in Figure~\ref{fig:ref80_dipole}d. This difference between parameterizations arises from the change in shape of the Fermi surface due to the $\gamma_4$ term, which, causes a shift-up in energy of the satellite Dirac cones (also seen in Figure \ref{fig:TB_vs_analytical}). As the value of $\gamma_4$ in the bilayer parametrization is three times larger than the one used for graphite, the shift up in energy of the satellite Dirac cones will be significantly larger. This means that the satellite Dirac cones will stop contributing to the BCD at higher electron densities. For the graphite parameterization, this shift up is much smaller and therefore the satellite Dirac cones will contribute up to lower electron densities.

When we neglect the $\gamma_4$-term a small negative peak appears in the Berry curvature dipole at very low electron densities when strain is applied in the y-direction, this is shown in panels \ref{fig:ref67_dipole}a-b and \ref{fig:ref80_dipole}a-b. In this case the satellite Dirac cones can reach energies below the central Dirac cone and below the gating voltage, which can be seen in Figure~\ref{fig:TB_vs_analytical}a-b. As a result, only the BC dipole density in the satellite Dirac will contribute to the BCD, and as this contribution will be negative, it will result in a small negative peak in the BCD.

These effects can be clearly observed in the BCD computed when using the TB model for both parameterizations. A sign change in the BCD can be observed for bilayer graphene strained in both the ZZ and AC directions. These sign changes are again linked to the detailed band-structure geometry in the Dirac cones. In the case of AC strain, a sign change occurs when the Fermi level falls below the wrapped satellite Dirac cones; for strain in the ZZ direction, the sign change occurs when the Fermi level drops below the saddle point. When the Fermi level reaches below these features part of the Dirac cone will no longer contribute to the BCD which causes the sign to change. In case of strain in the AC direction only the warped main cone will contribute; for strain in the ZZ direction only the warped satellite cones contribute. As the position of the satellite Dirac cones changes with strain, so will the point at which a sign change occurs. These sign changes are more pronounced when the parameterization for bilayer graphene is used due to the larger $\gamma_4$ term, which causes a larger upward shift of the satellite Dirac cones particularly for the TB-model where this shift is again larger as when the analytical model is used. The larger upward shift causes the satellite cones to stop contributing at higher electron densities, making the resulting sign changes more significant. This is illustrated by the contour lines in Figure~\ref{fig:ref67_berry} and for higher strains in Figure~\ref{fig:Berry_moments} which show the shape of the Fermi surface at low energies and the first moment of the BC.
\begin{figure*}[!th]
\centering
\includegraphics[width=\figwidth]{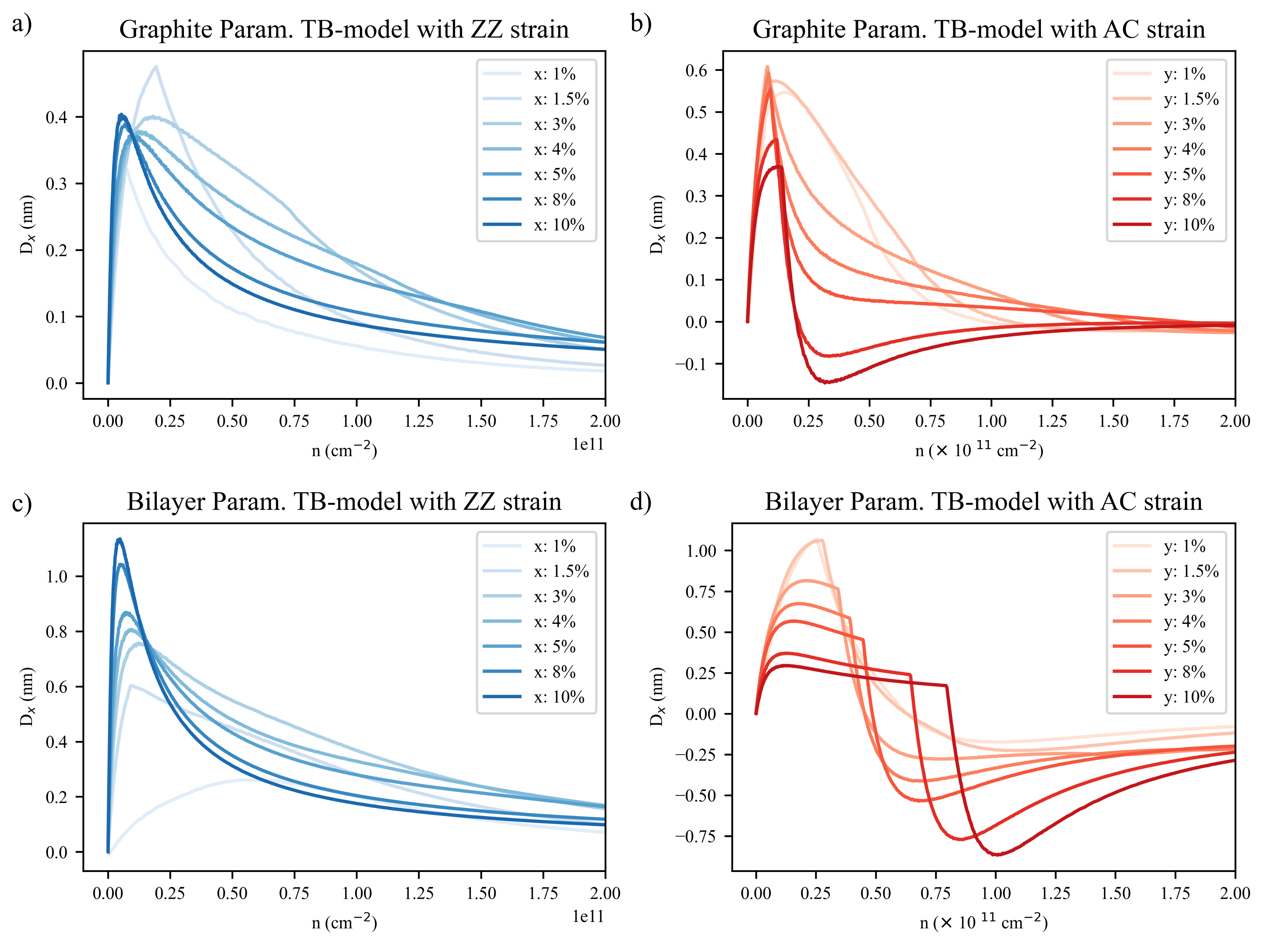}
\includegraphics[width=\figwidth]{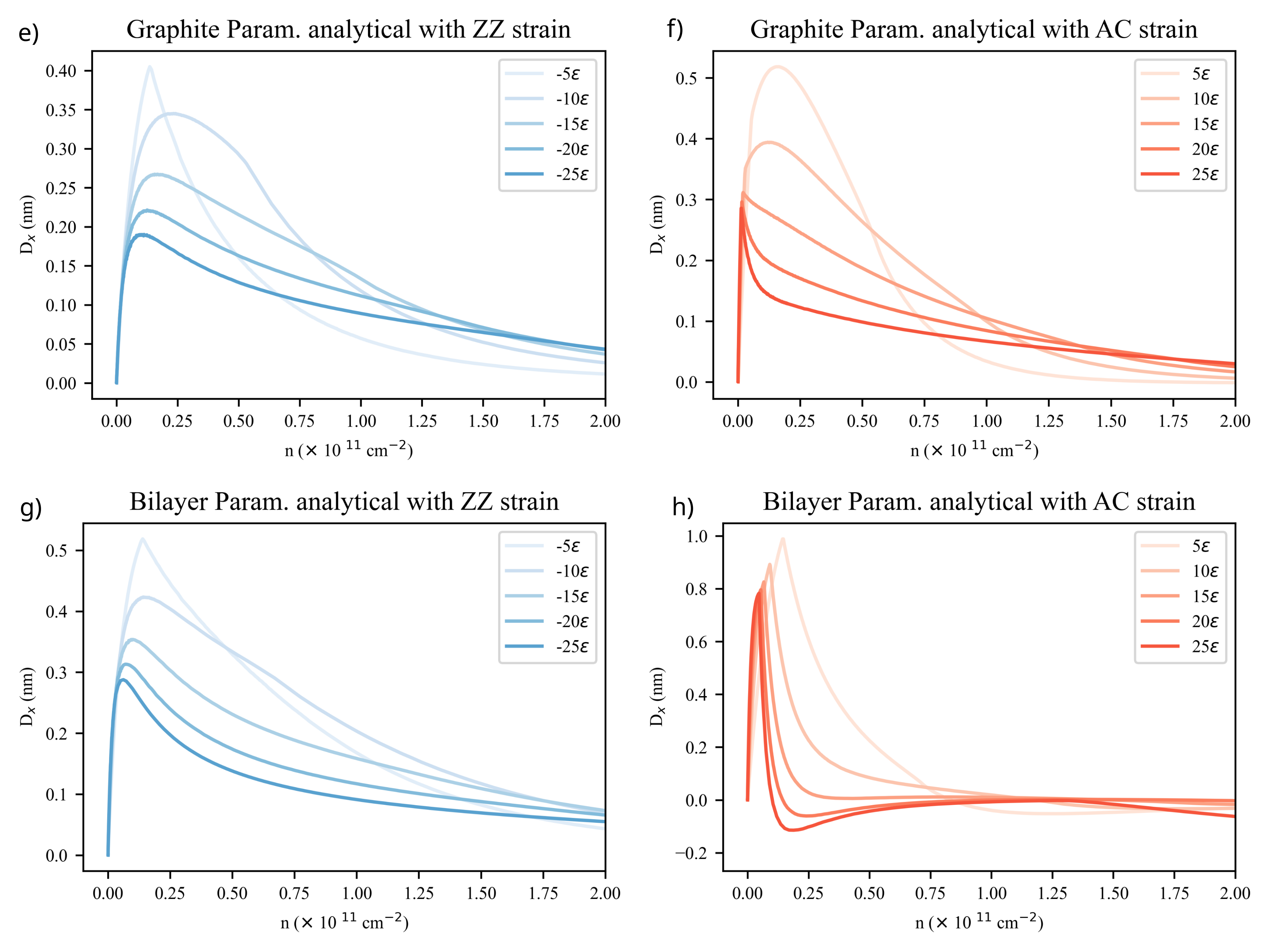}
\caption{\label{fig:high_strain} The BCD as a function of the electron density calculated for large strains: a)-d) TB model for ZZ/AC strain for graphite/bilayer parametrizations, e)-h) analytical model for ZZ/AC strain for graphite/bilayer parametrizations. Here the gating potential is U = 10 meV.
}
\end{figure*}

\subsection{The Berry curvature dipole at higher strains}

We further take advantage of the extended range in the accuracy of the TB model and perform calculations of the BCD of bilayer graphene under higher strains. Figure~\ref{fig:high_strain}a-d shows how the BCD changes with higher strains along ZZ and AC directions. To keep the calculations experimentally relevant we considered strains only up to 10\%. We can notice that the BCD does not simply increase with strain but behaves non-monotonically, depending on where the Fermi energy is located. For strain in the ZZ direction we find that the BCD at very low electron densities ($n < 2.5 \times 10^{10} cm^{-2}$) the BCD will increase with strain. At higher electron densities the BCD will first increase and then decrease with strain. For strain in the AC direction, we find that the BCD will decrease with strain at very low electron densities, while at high electron densities, the BCD will even change sign.

To evaluate the accuracy of the continuum model we show in Figure~\ref{fig:high_strain}e-h how higher strain affects the BCD according to the analytical model with both parameterizations. The effect strain has on the BCD is significantly different from what we obtained when using the TB model. Firstly, at high strains in the ZZ direction, by adding a negative strain term, the peak BCD now decreases with increased strain. This is the case for both parameterizations. At high strain in the AC direction, by adding a positive strain term, the BCD again decreases with strain. Furthermore the BCD does not change sign when using the graphite model. The sign change in the BCD occurs for the bilayer graphene model only at very high strain, but the absolute value is much smaller than the value predicted by the TB approach.

While the TB model is expected to remain accurate in the high strain regime, we see that analytical approach in calculating the BCD already breaks down. This can be clearly seem from our work, which shows that both approaches give similar BCD at low strain, but we find that there are significant differences at higher strains. 
\begin{figure*}[!t]
\centering
\includegraphics[width=\figwidth]{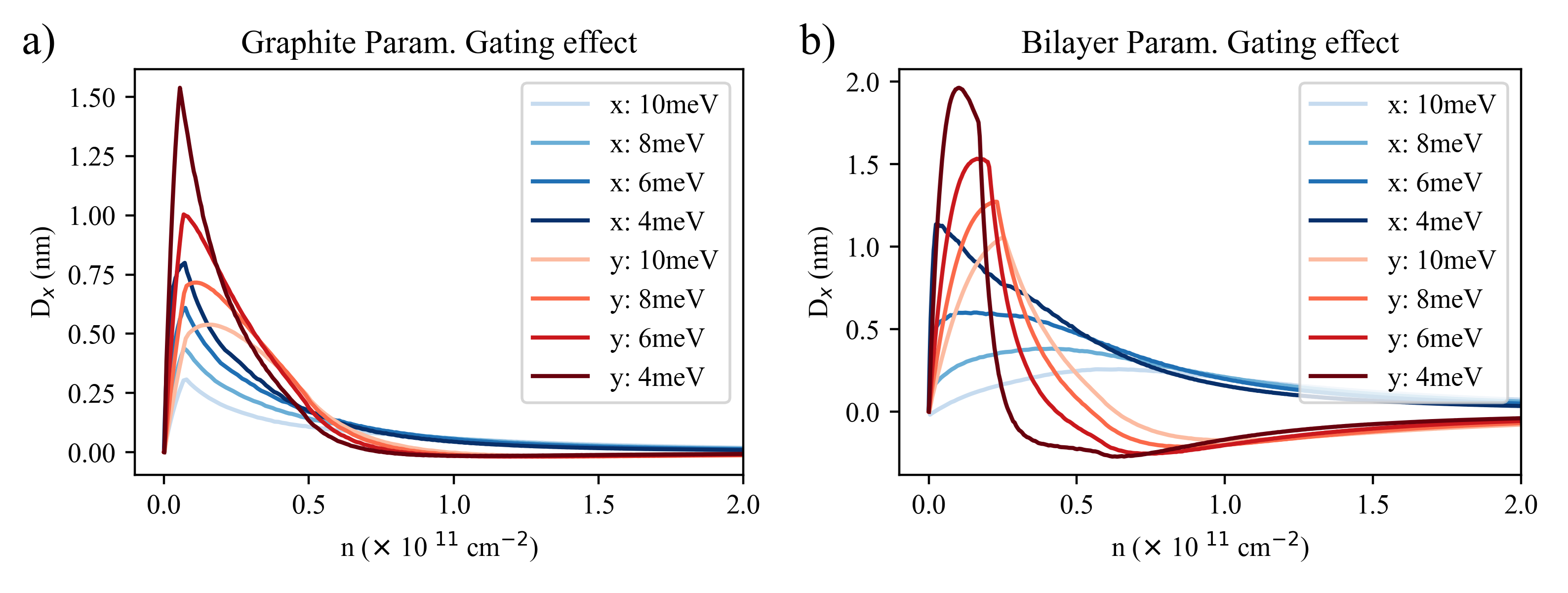}
\caption{\label{fig:Gating_plot} Effect of interlayer gating potentials on the BCD as a function of the electron density for bilayer graphene with a strain of 1\% in either the ZZ (x) or AC (y) direction. a) considering bulk graphite parametrization, and b) bilayer graphene parametrization.}
\end{figure*}

\subsection{Changing the interlayer gating potential}

We have so far fixed the gating potential at $U=10meV$, but in reality the shape of the dispersion and the BC depend on the strength of the gating potential. To evaluate its effect on the BCD we show in Figure~\ref{fig:Gating_plot} the evolution of the BCD as a function of the electron density as we change the gating potential. Here we use the TB model and 1\% strain in the ZZ or AC directions. We notice that the magnitude of the peak in the BCD increases as the magnitude of the gating potential decreases. This is because the BC scales inversely with the size in the band gap formed by the gating potential. Therefore, lowering the gating potential will increase the BC. On the other hand, as the gating potential is reduced, the BC becomes highly peaked near the band edges, resulting in a faster decay of the BCD as a function of doping. It is interesting to note that for strains in the AC direction, for a certain range of strains and doping, the interlayer gating potential can be used to change the sign of the BCD. For example, at $\%1$ strain and $n=0.5 \times 10^{-11} cm^{-2}$ the sign of the BCD changes when tuning the interlayer gating potential from 4meV to 10meV.

\begin{figure*}[!t]
\centering
\includegraphics[width=\figwidth]{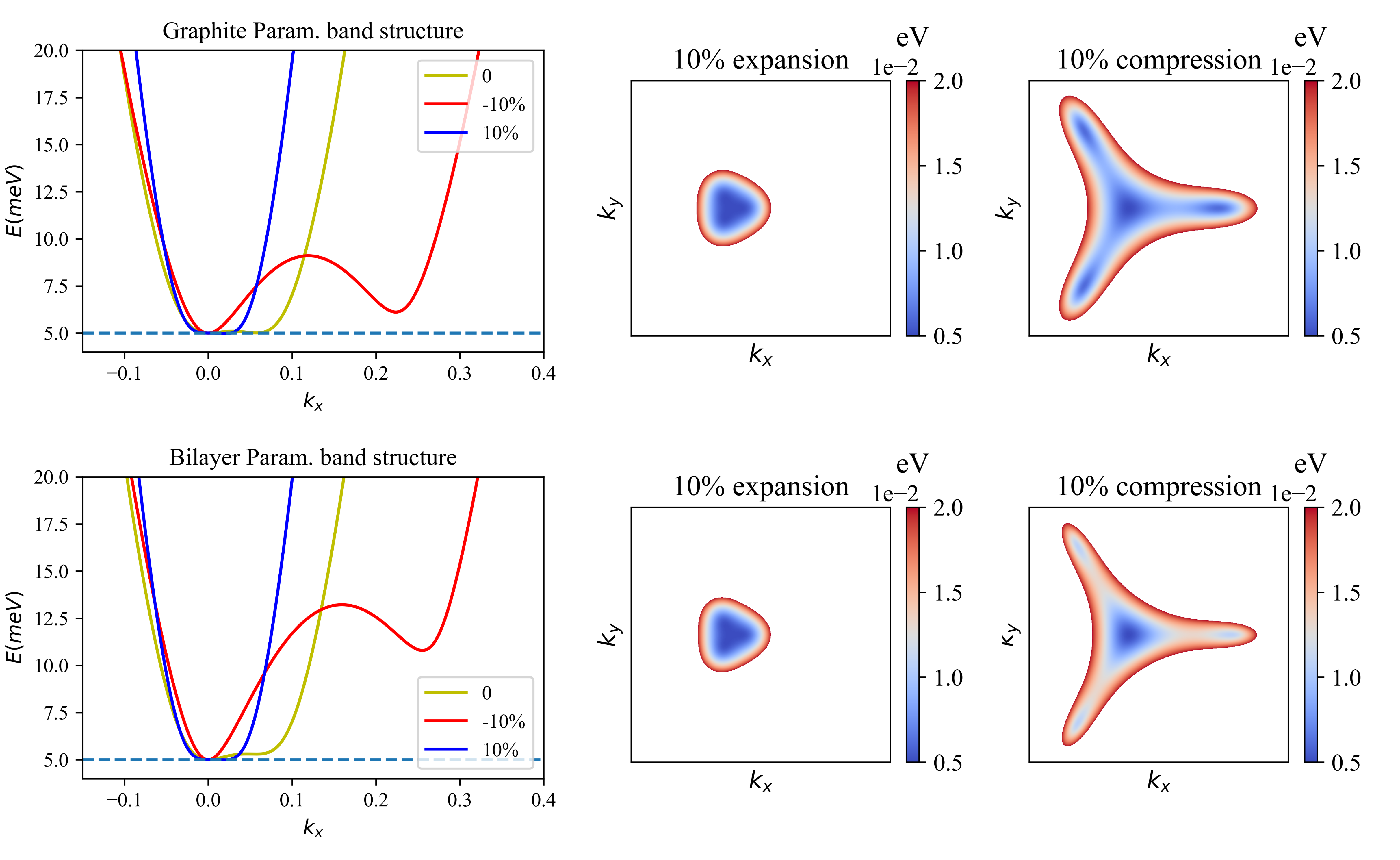}
\caption{\label{fig:comp_exp} Dispersion of bilayer graphene as a function of interlayer distance. Left panels show a slice of the Dirac cone along the x-axis while middle and right panels show a top-down view of the dispersion. Top row corresponds to graphite parameters, while bottom row corresponds to graphene parameters. Here the gating potential is U = 10 meV. }
\end{figure*}

\begin{figure*}[t]
\centering
\includegraphics[width=\figwidth]{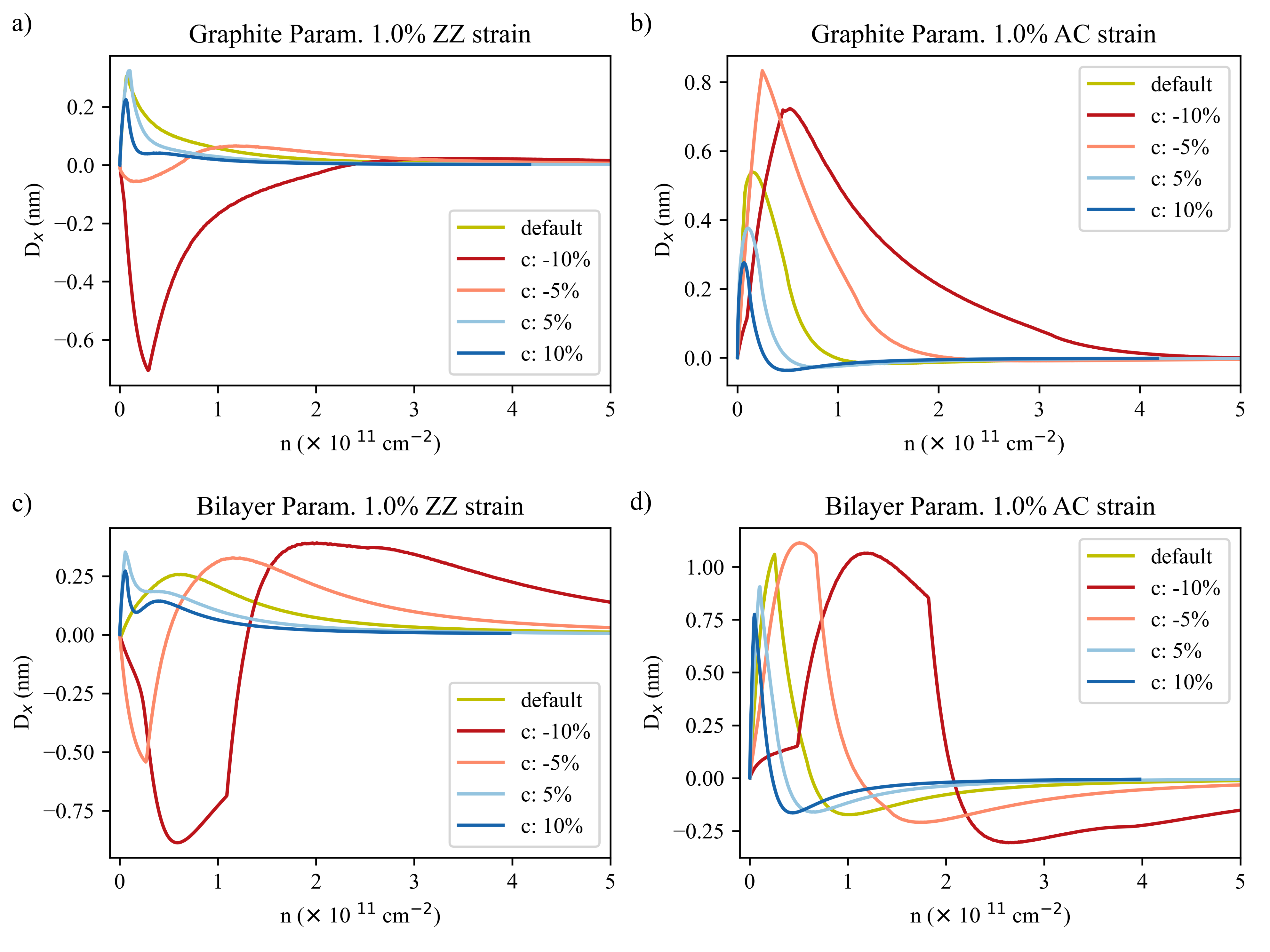}
\caption{\label{fig:compression_dipole} The BCD calculated using the TB model as a function of the electron density, for various interlayer distances and an added 1\% in-plane strain: a)-b) BCD for graphite parameters and ZZ/AC strain, c)-d) BCD for bilayer graphene parameters and ZZ/AC strain. Here the gating potential is U = 10 meV.}
\end{figure*}

\subsection{Changing the interlayer distance}

In this section we consider another experimentally relevant tuning parameter, the out-of-plane strain. This will affect the interlayer distance, which will change all interlayer hoppings while not affecting the $\gamma_0$ intra-layer hopping term. Figure~\ref{fig:comp_exp} shows how out-of-plane expansion or compression affects the band structure of models graphene/graphite parametrizations. We find that decreasing the interlayer distance (negative strain percentage) will cause the interlayer hopping amplitudes to become larger, which will in turn cause the Dirac cones to widen when the layers are compressed. Increasing the distance between the layers will decrease the hopping amplitudes and cause the Dirac cone to narrow. Figure~\ref{fig:comp_exp} also shows that, as expected, changing the interlayer distance will not break the trigonal symmetry of the Dirac cone. The BCD will remain zero unless an in-plane strain is applied to the graphene lattice as well.

Figure~\ref{fig:compression_dipole} shows that when an in-plane strain is combined with a change in the interlayer distance, the resulting broadening of the Dirac cone leads to a substantial enhancement of the BCD. In contrast, increasing the interlayer distance causes a narrowing of the cone, which reduces the magnitude of the BCD. This enhancement of the BCD under layer compression is particularly relevant for experiments, as it shows that applying out-of-plane pressure to compress the layers would strengthen the anomalous Hall response, thereby making it easier to detect experimentally.

\section{Conclusion}
The BCD in strained bilayer graphene proves sensitive to details of the electronic structure that are often treated as secondary. Our TB calculations reveal that the choice between  bilayer graphene versus bulk graphite parameters leads to qualitatively different predictions for the BCD as a function of doping, particularly sign changes that occur at experimentally accessible carrier densities. We find that this difference arises from the $\gamma_4$ skew hopping, which shifts the satellite Dirac cones in energy and thus controls when they contribute to the BCD. 

When comparing the TB and continuum approaches, we find reasonable agreement at strains below roughly 1\%, but the effective Hamiltonian description break down at larger deformations where the TB model is expected to remains reliable. At 10\% strain, the two methods predict not only different magnitudes but different trends with increasing strain. The interplay between in-plane and out-of-plane strain offers additional tunability. Layer compression enhances all interlayer hoppings, broadening the Dirac cones and amplifying the BCD without breaking the trigonal symmetry that in-plane strain disrupts. Combining moderate uniaxial strain with out-of-plane pressure could therefore provide a practical route to maximizing the nonlinear Hall response in experiments.

Several questions remain open. We have neglected the Poisson ratio of graphene, treating $\epsilon_x$ and $\epsilon_y$ as independent. Incorporating realistic elastic coupling would modify the quantitative predictions. Temperature-dependent effects or disorder scattering also warrant investigation. Nonetheless, the present results establish that realistic TB modeling is essential for reliable predictions of strain-tuned BC phenomena in bilayer graphene.

\clearpage
\bibliography{BCD_paper}

@article{xiao_berry_2010,
	title = {Berry phase effects on electronic properties},
	volume = {82},
	copyright = {http://link.aps.org/licenses/aps-default-license},
	issn = {0034-6861, 1539-0756},
	url = {https://link.aps.org/doi/10.1103/RevModPhys.82.1959},
	doi = {10.1103/RevModPhys.82.1959},
	language = {en},
	number = {3},
	urldate = {2025-04-11},
	journal = {Reviews of Modern Physics},
	author = {Xiao, Di and Chang, Ming-Che and Niu, Qian},
	month = jul,
	year = {2010},
	pages = {1959--2007},
}

@article{ho_zero-magnetic-field_2021,
	title = {Zero-magnetic-field {Hall} effects in artificially corrugated bilayer graphene},
	volume = {4},
	issn = {2520-1131},
	url = {http://arxiv.org/abs/1910.07509},
	doi = {10.1038/s41928-021-00537-5},
	language = {en},
	number = {2},
	urldate = {2025-04-11},
	journal = {Nature Electronics},
	author = {Ho, Sheng-Chin and Chang, Ching-Hao and Hsieh, Yu-Chiang and Lo, Shun-Tsung and Huang, Botsz and Vu, Thi-Hai-Yen and Ortix, Carmine and Chen, Tse-Ming},
	month = feb,
	year = {2021},
	note = {arXiv:1910.07509 [cond-mat]},
	pages = {116--125},
}

@article{battilomo_berry_2019,
	title = {Berry {Curvature} {Dipole} in {Strained} {Graphene}: a {Fermi} {Surface} {Warping} {Effect}},
	volume = {123},
	issn = {0031-9007, 1079-7114},
	shorttitle = {Berry {Curvature} {Dipole} in {Strained} {Graphene}},
	url = {http://arxiv.org/abs/1910.09872},
	doi = {10.1103/PhysRevLett.123.196403},
	language = {en},
	number = {19},
	urldate = {2025-04-11},
	journal = {Physical Review Letters},
	author = {Battilomo, Raffaele and Scopigno, Niccolo' and Ortix, Carmine},
	month = nov,
	year = {2019},
	note = {arXiv:1910.09872 [cond-mat]},
	pages = {196403},
}

@article{bandyopadhyay_non-linear_2024,
	title = {Non-linear {Hall} {Effects}: {Mechanisms} and {Materials}},
	volume = {8},
	issn = {27729494},
	shorttitle = {Non-linear {Hall} {Effects}},
	url = {http://arxiv.org/abs/2401.02282},
	doi = {10.1016/j.mtelec.2024.100101},
	language = {en},
	urldate = {2025-04-11},
	journal = {Materials Today Electronics},
	author = {Bandyopadhyay, Arka and Joseph, Nesta Benno and Narayan, Awadhesh},
	month = jun,
	year = {2024},
	note = {arXiv:2401.02282 [cond-mat]},
	pages = {100101},
}

@article{pereira_tight-binding_2009,
	title = {Tight-binding approach to uniaxial strain in graphene},
	volume = {80},
	copyright = {http://link.aps.org/licenses/aps-default-license},
	issn = {1098-0121, 1550-235X},
	url = {https://link.aps.org/doi/10.1103/PhysRevB.80.045401},
	doi = {10.1103/PhysRevB.80.045401},
	language = {en},
	number = {4},
	urldate = {2025-04-11},
	journal = {Physical Review B},
	author = {Pereira, Vitor M. and Castro Neto, A. H. and Peres, N. M. R.},
	month = jul,
	year = {2009},
	pages = {045401},
}

@article{slater_simplified_1954,
	title = {Simplified {LCAO} {Method} for the {Periodic} {Potential} {Problem}},
	volume = {94},
	copyright = {http://link.aps.org/licenses/aps-default-license},
	issn = {0031-899X},
	url = {https://link.aps.org/doi/10.1103/PhysRev.94.1498},
	doi = {10.1103/PhysRev.94.1498},
	language = {en},
	number = {6},
	urldate = {2025-04-11},
	journal = {Physical Review},
	author = {Slater, J. C. and Koster, G. F.},
	month = jun,
	year = {1954},
	pages = {1498--1524},
}

@article{mccann_electronic_2013,
	title = {The electronic properties of bilayer graphene},
	volume = {76},
	issn = {0034-4885, 1361-6633},
	url = {http://arxiv.org/abs/1205.6953},
	doi = {10.1088/0034-4885/76/5/056503},
	language = {en},
	number = {5},
	urldate = {2025-04-11},
	journal = {Reports on Progress in Physics},
	author = {McCann, Edward and Koshino, Mikito},
	month = may,
	year = {2013},
	note = {arXiv:1205.6953 [cond-mat]},
	pages = {056503},
}

@article{mucha-kruczynski_landau_2011,
	title = {Landau levels in deformed bilayer graphene at low magnetic fields},
	volume = {151},
	issn = {00381098},
	url = {http://arxiv.org/abs/1109.3348},
	doi = {10.1016/j.ssc.2011.05.019},
	language = {en},
	number = {16},
	urldate = {2025-04-11},
	journal = {Solid State Communications},
	author = {Mucha-Kruczynski, M. and Aleiner, I. L. and Fal'ko, V. I.},
	month = aug,
	year = {2011},
	note = {arXiv:1109.3348 [cond-mat]},
	pages = {1088--1093},
}

@article{vozmediano_gauge_2010,
	title = {Gauge fields in graphene},
	volume = {496},
	issn = {03701573},
	url = {http://arxiv.org/abs/1003.5179},
	doi = {10.1016/j.physrep.2010.07.003},
	language = {en},
	number = {4-5},
	urldate = {2025-04-11},
	journal = {Physics Reports},
	author = {Vozmediano, M. A. H. and Katsnelson, M. I. and Guinea, F.},
	month = nov,
	year = {2010},
	note = {arXiv:1003.5179 [cond-mat]},
	pages = {109--148},
}

@article{dresselhaus_intercalation_2002,
	title = {Intercalation compounds of graphite},
	volume = {51},
	issn = {0001-8732, 1460-6976},
	url = {http://www.tandfonline.com/doi/abs/10.1080/00018730110113644},
	doi = {10.1080/00018730110113644},
	language = {en},
	number = {1},
	urldate = {2025-04-11},
	journal = {Advances in Physics},
	author = {Dresselhaus, M. S. and Dresselhaus, G.},
	month = jan,
	year = {2002},
	pages = {1--186},
}

@article{kuzmenko_determination_2009,
	title = {Determination of the gate-tunable band gap and tight-binding parameters in bilayer graphene using infrared spectroscopy},
	volume = {80},
	copyright = {http://link.aps.org/licenses/aps-default-license},
	issn = {1098-0121, 1550-235X},
	url = {https://link.aps.org/doi/10.1103/PhysRevB.80.165406},
	doi = {10.1103/PhysRevB.80.165406},
	language = {en},
	number = {16},
	urldate = {2025-04-11},
	journal = {Physical Review B},
	author = {Kuzmenko, A. B. and Crassee, I. and Van Der Marel, D. and Blake, P. and Novoselov, K. S.},
	month = oct,
	year = {2009},
	pages = {165406},
}

@article{moon_energy_2012,
	title = {Energy spectrum and quantum {Hall} effect in twisted bilayer graphene},
	volume = {85},
	copyright = {http://link.aps.org/licenses/aps-default-license},
	issn = {1098-0121, 1550-235X},
	url = {https://link.aps.org/doi/10.1103/PhysRevB.85.195458},
	doi = {10.1103/PhysRevB.85.195458},
	language = {en},
	number = {19},
	urldate = {2025-04-14},
	journal = {Physical Review B},
	author = {Moon, Pilkyung and Koshino, Mikito},
	month = may,
	year = {2012},
	pages = {195458},
}

@misc{moldovan_pybinding_2020,
	title = {pybinding v0.9.5: a {Python} package for tight-binding calculations},
	shorttitle = {pybinding v0.9.5},
	url = {https://zenodo.org/records/4010216},
	urldate = {2025-04-14},
	publisher = {Zenodo},
	author = {Moldovan, Dean and Anđelković, Miša and Peeters, Francois},
	month = aug,
	year = {2020},
	doi = {10.5281/zenodo.4010216},
}

@article{virtanen_scipy_2020,
	title = {{SciPy} 1.0: fundamental algorithms for scientific computing in {Python}},
	volume = {17},
	copyright = {2020 The Author(s)},
	issn = {1548-7105},
	shorttitle = {{SciPy} 1.0},
	url = {https://www.nature.com/articles/s41592-019-0686-2},
	doi = {10.1038/s41592-019-0686-2},
	language = {en},
	number = {3},
	urldate = {2025-04-14},
	journal = {Nature Methods},
	author = {Virtanen, Pauli and Gommers, Ralf and Oliphant, Travis E. and Haberland, Matt and Reddy, Tyler and Cournapeau, David and Burovski, Evgeni and Peterson, Pearu and Weckesser, Warren and Bright, Jonathan and van der Walt, Stéfan J. and Brett, Matthew and Wilson, Joshua and Millman, K. Jarrod and Mayorov, Nikolay and Nelson, Andrew R. J. and Jones, Eric and Kern, Robert and Larson, Eric and Carey, C. J. and Polat, İlhan and Feng, Yu and Moore, Eric W. and VanderPlas, Jake and Laxalde, Denis and Perktold, Josef and Cimrman, Robert and Henriksen, Ian and Quintero, E. A. and Harris, Charles R. and Archibald, Anne M. and Ribeiro, Antônio H. and Pedregosa, Fabian and van Mulbregt, Paul},
	month = mar,
	year = {2020},
	note = {Publisher: Nature Publishing Group},
	pages = {261--272},
}

@article{fukui_chern_2005,
	title = {Chern Numbers in Discretized Brillouin Zone: Efficient Method of Computing (Spin) Hall Conductances},
	volume = {74},
	issn = {0031-9015, 1347-4073},
	url = {http://arxiv.org/abs/cond-mat/0503172},
	doi = {10.1143/JPSJ.74.1674},
	shorttitle = {Chern Numbers in Discretized Brillouin Zone},
	pages = {1674--1677},
	number = {6},
	journal = {Journal of the Physical Society of Japan},
	shortjournal = {J. Phys. Soc. Jpn.},
	author = {Fukui, Takahiro and Hatsugai, Yasuhiro and Suzuki, Hiroshi},
	year = {2025},
	date = {2005-06},
	langid = {english},
}

@article{du_quantum_2021,
	title = {Quantum theory of the nonlinear Hall effect},
	volume = {12},
	rights = {2021 The Author(s)},
	issn = {2041-1723},
	url = {https://www.nature.com/articles/s41467-021-25273-4},
	doi = {10.1038/s41467-021-25273-4},
	pages = {5038},
	number = {1},
	journal = {Nature Communications},
	shortjournal = {Nat Commun},
	author = {Du, Z. Z. and Wang, C. M. and Sun, Hai-Peng and Lu, Hai-Zhou and Xie, X. C.},
	urldate = {2025-12-17},
	year = {2021},
	langid = {english},
	note = {Publisher: Nature Publishing Group},
}

@article{kang_nonlinear_2019,
	title = {Nonlinear anomalous Hall effect in few-layer {WTe}2},
	volume = {18},
	issn = {1476-4660},
	doi = {10.1038/s41563-019-0294-7},
	pages = {324--328},
	number = {4},
	journal = {Nature Materials},
	shortjournal = {Nat Mater},
	author = {Kang, Kaifei and Li, Tingxin and Sohn, Egon and Shan, Jie and Mak, Kin Fai},
	year = {2019},
	pmid = {30804510},
}

@article{xiong_tunable_2025,
	title = {Tunable linear and nonlinear anomalous {Hall} transport in two-dimensional {CrPS}$_{\textrm{4}}$},
	volume = {37},
	issn = {0953-8984, 1361-648X},
	url = {https://iopscience.iop.org/article/10.1088/1361-648X/ae084d},
	doi = {10.1088/1361-648X/ae084d},
	number = {41},
	urldate = {2025-12-19},
	journal = {Journal of Physics: Condensed Matter},
	author = {Xiong, Lulu and Cao, Jin and Yang, Fan and Yang, Xiaoxin and Lai, Shen and Sheng, Xian-Lei and Xiao, Cong and Yang, Shengyuan A},
	month = oct,
	year = {2025},
	pages = {415501},
}

@article{wang_nonlinear_2025,
	title = {Nonlinear {Hall} effect in two-dimensional materials},
	volume = {5},
	issn = {2770-2995, 2770-2995},
	url = {https://www.oaepublish.com/articles/microstructures.2024.129},
	doi = {10.20517/microstructures.2024.129},
	number = {3},
	urldate = {2025-12-19},
	journal = {Microstructures},
	author = {Wang, Shuo and Niu, Wei and Fang, Yue-Wen},
	month = apr,
	year = {2025},
}

@article{qin_strain_2021,
	title = {Strain {Tunable} {Berry} {Curvature} {Dipole}, {Orbital} {Magnetization} and {Nonlinear} {Hall} {Effect} in {WSe}$_{\textrm{2}}$ {Monolayer}*},
	volume = {38},
	issn = {0256-307X, 1741-3540},
	url = {https://iopscience.iop.org/article/10.1088/0256-307X/38/1/017301},
	doi = {10.1088/0256-307X/38/1/017301},
	number = {1},
	urldate = {2025-12-19},
	journal = {Chinese Physics Letters},
	author = {Qin, Mao-Sen and Zhu, Peng-Fei and Ye, Xing-Guo and Xu, Wen-Zheng and Song, Zhen-Hao and Liang, Jing and Liu, Kaihui and Liao, Zhi-Min},
	month = jan,
	year = {2021},
	pages = {017301},
}

@article{duan_giant_2022,
	title = {Giant {Second}-{Order} {Nonlinear} {Hall} {Effect} in {Twisted} {Bilayer} {Graphene}},
	volume = {129},
	issn = {0031-9007, 1079-7114},
	url = {https://link.aps.org/doi/10.1103/PhysRevLett.129.186801},
	doi = {10.1103/PhysRevLett.129.186801},
	language = {en},
	number = {18},
	urldate = {2025-12-19},
	journal = {Physical Review Letters},
	author = {Duan, Junxi and Jian, Yu and Gao, Yang and Peng, Huimin and Zhong, Jinrui and Feng, Qi and Mao, Jinhai and Yao, Yugui},
	month = oct,
	year = {2022},
	pages = {186801},
}

@article{he_graphene_2022,
	title = {Graphene moiré superlattices with giant quantum nonlinearity of chiral {Bloch} electrons},
	volume = {17},
	issn = {1748-3387, 1748-3395},
	url = {https://www.nature.com/articles/s41565-021-01060-6},
	doi = {10.1038/s41565-021-01060-6},
	language = {en},
	number = {4},
	urldate = {2025-12-19},
	journal = {Nature Nanotechnology},
	author = {He, Pan and Koon, Gavin Kok Wai and Isobe, Hiroki and Tan, Jun You and Hu, Junxiong and Neto, Antonio H. Castro and Fu, Liang and Yang, Hyunsoo},
	month = apr,
	year = {2022},
	pages = {378--383},
}

@article{li_new_2014,
	title = {New position of Dirac points in the strained graphene reciprocal lattice},
	volume = {4},
	issn = {2158-3226},
	url = {https://doi.org/10.1063/1.4893239},
	doi = {10.1063/1.4893239},
	pages = {087119},
	number = {8},
	journal = {{AIP} Advances},
	author = {Li, Cui-Lian},
	urldate = {2025-12-30},
	date = {2014-08-13},
    year = {2014},
}

\newpage
\section{Supplementary figures}
\setcounter{equation}{0}
\setcounter{figure}{0}
\setcounter{table}{0}
\setcounter{page}{1}
\renewcommand{\thepage}{S\arabic{page}}
\renewcommand{\thefigure}{S\arabic{figure}}

\begin{figure*}[th]
\centering
\includegraphics[width=0.55\figwidth]{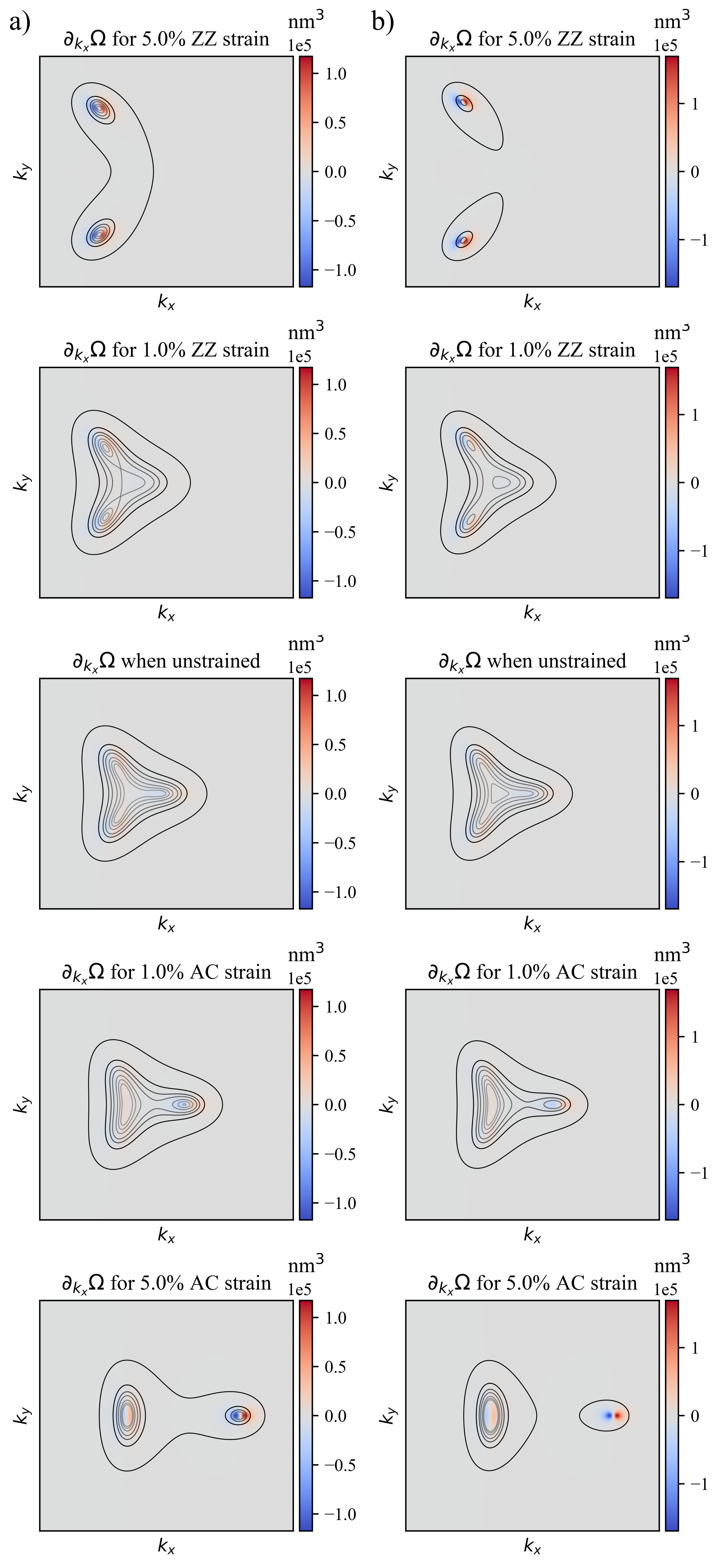}
\caption{\label{fig:Berry_moments} The first moment of the BC and contours showing the Fermi surface for various energies calculated using the TB model using either a) bulk graphite or b) bilayer graphene parametrizations. Here the gating potential is U = 10 meV.}
\end{figure*}

\begin{figure*}[ht]
\centering
\includegraphics[width=\figwidth]{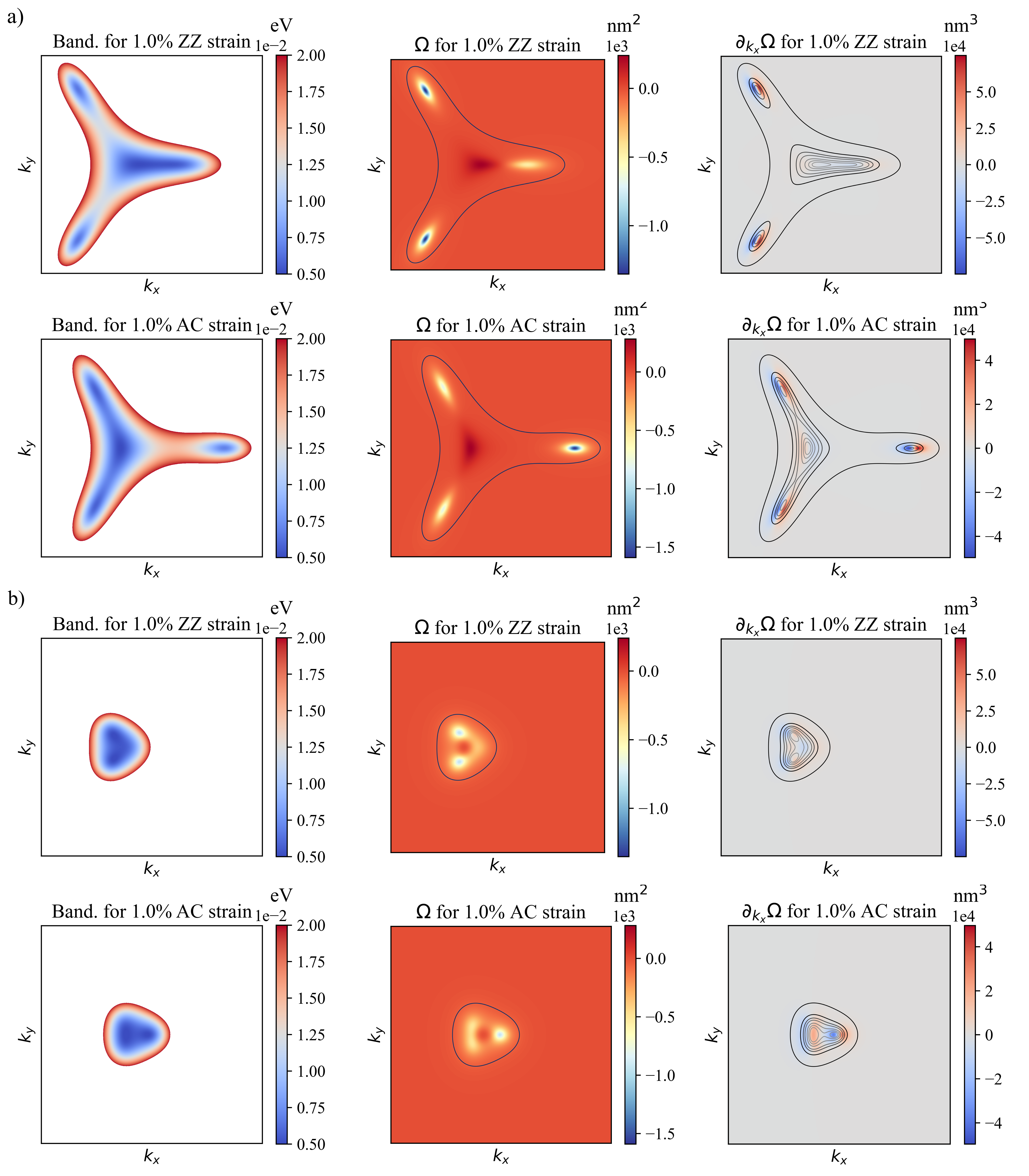}
\caption{\label{fig:compression_ref67} The dispersion, the BC, and the first derivative of the BC in the low energy region of the Dirac cone for various interlayer distances.  The black lines show the shape of the Fermi surface. a) interlayer distance is decreased by 10\%, and figure b) interlayer distance is increased by 10\%. Here the gating potential is U = 10 meV and bilayer graphene parametrization is used.}
\end{figure*}

\end{document}